\documentclass{iau} 
\usepackage{graphicx}
\usepackage{natbib}
\usepackage{subfigure}
\usepackage{wrapfig}

\title[Magnetospheres of B-type Stars] 
{Investigating the Magnetospheres of Rapidly Rotating B-type Stars}

\author[C. L. Fletcher]   
{C. L. Fletcher$^1$, 
 V. Petit$^2$, Y. Naz\'e$^3$, G. A. Wade$^4$, R. H. Townsend$^5$, S. P. Owocki$^2$, D. H. Cohen$^6$, A. David-Uraz$^2$, M. Shultz$^7$ }

\affiliation{$^1$Department of Physics and Space Sciences, Florida Institute of Technology, Melbourne, FL, 32904, USA, cfletcher2013@my.fit.edu  \\[\affilskip]
$^2$ Department of Physics and Astronomy, Bartol Research Institute, University of Delaware, Newark, DE 19716, USA \\[\affilskip]
$^3$ FNRS GAPHE - STAR - Institut d’Astrophysique et de G\'eophysique (B5C), Universit\'e de Li\`ege, All\'ee du 6 Ao\^ut 19c, 4000-Li\'ege, Belgium \\[\affilskip]
$^4$Department of Physics, Royal Military College of Canada, PO Box 17000 Station Forces, Kingston, ON, Canada K7K 0C6 \\[\affilskip]
$^5$ Department of Astronomy, University of Wisconsin-Madison, 5534 Sterling Hall, 475 N Charter Street, Madison, WI 53706, USA \\[\affilskip]
$^6$Department of Physics and Astronomy, Swarthmore College, 500 College Ave., Swarthmore, PA 19081, USA \\[\affilskip]
$^7$Department of Physics and Astronomy, Uppsala University, Box 516, Uppsala 75120, Sweden}

\pubyear{2017}
\volume{329}  
\jname{The Lives and Death Throes of Massive Stars}
\editors{A.C. Editor, B.D. Editor \& C.E. Editor, eds.}
\begin{document}

\maketitle

\begin{abstract}
Recent spectropolarimetric surveys of bright, hot stars have found that $\sim 10\%$ of OB-type stars contain strong (mostly dipolar) surface magnetic fields ($\sim$kG).  The prominent paradigm describing the interaction between the stellar winds and the surface magnetic field is the magnetically confined wind shock (MCWS) model.  In this model, the stellar wind plasma is forced to move along the closed field loops of the magnetic field, colliding at the magnetic equator, and creating a shock.  As the shocked material cools radiatively it will emit X-rays. Therefore, X-ray spectroscopy is a key tool in detecting and characterizing the hot wind material confined by the magnetic fields of these stars.  Some B-type stars are found to have very short rotational periods.  The effects of the rapid rotation on the X-ray production within the magnetosphere have yet to be explored in detail. The added centrifugal force due to rapid rotation is predicted to cause faster wind outflows along the field lines, leading to higher shock temperatures and harder X-rays. However, this is not observed in all rapidly rotating magnetic B-type stars.  In order to address this from a theoretical point of view, we use the X-ray Analytical Dynamical Magnetosphere (XADM) model, originally developed for slow rotators, with an implementation of new rapid rotational physics. Using X-ray spectroscopy from ESA’s XMM-Newton space telescope, we observed 5 rapidly rotating B-types stars to add to the previous list of observations.  Comparing the observed X-ray luminosity and hardness ratio to that predicted by the XADM allows us to determine the role the added centrifugal force plays in the magnetospheric X-ray emission of these stars. 
\end{abstract}

\firstsection 

\section{Characterizing the Magnetic Fields of Massive Stars}
Recent surveys have leveraged the development of improved spectropolarimeters to detect and characterize magnetic fields in a large sample of OB-type stars \citep{mimes, fossati16}. These studies have shown that $\sim 10\%$ of massive stars host surface magnetic fields that are strong ($\sim 1$kG) and mostly dipolar. The stellar winds of massive stars are on the order of the stellar effective temperature and photoionized causing them to interact with the magnetic field. \\
\indent The Magnetically Confined Wind Shock (MCWS) model presented by \cite{babel} has become the central idea explaining the X-ray emission of magnetic OB stars. In this scenario, the stellar wind plasma is forced to move along the magnetic field lines. In regions near the star, where the magnetic field lines are closed loops, the stellar wind channeled from both footpoints (located in the northern and southern hemispheres) will collide at the magnetic equator. Since the stellar wind is supersonic, the collision creates a shock with high post shock temperatures ($\sim1-10$ MK). The shocked material then cools radiatively, emitting X-ray photons \citep{ud-doula}. Therefore, X-ray observations are a key tool to probe these magnetospheres.
\begin{wrapfigure}{r}{0.5\textwidth}
\includegraphics[width=0.49\textwidth]{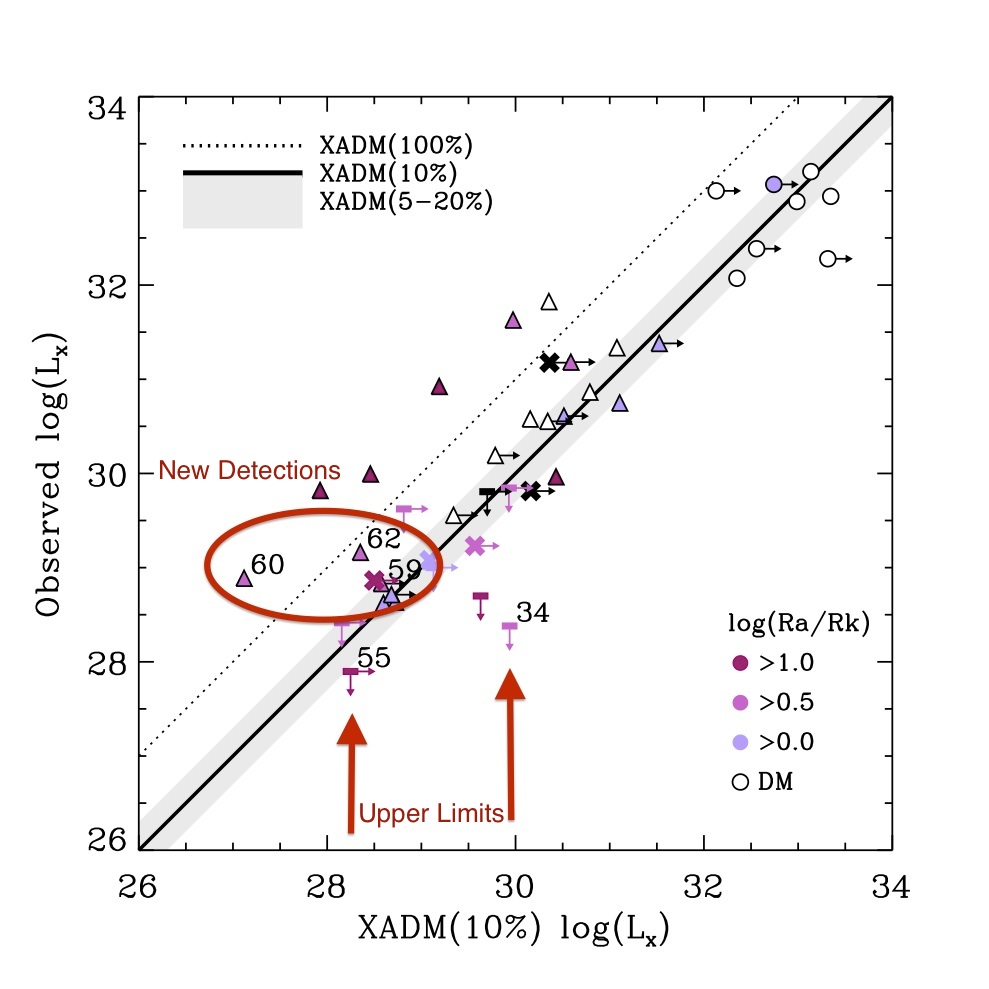}
\caption{The predicted versus observed X-ray luminosity plot from \cite{xpod} updated with new X-ray observations of 5 rapidly rotating B-type stars observed with XMM-Newton. The three new sources that have detections are shown with the numbered triangles and the two upper limits are the numbered rectangles with arrows. The numbers are the identification number taken from \cite{ipod} with the B-type stars in triangles, the O-type stars in circles and the undetected sources in rectangles. The color scheme corresponds to the size of the CM with the darker the color having a larger CM and the white only having a DM.  \label{fig:newdetect}}
\end{wrapfigure}
\indent An analytical approach to modeling the confined material in stars with slow rotation, called the Analytic Dynamical Magnetosphere (ADM) model, was developed by \cite{adm}. This model simplifies the complex results from previous MHD simulations \citep{ud-doula,ud-doula2,ud-doula3} by assuming a time averaged view of the processes in the confined material. \cite{xadm} derived predictions for X-ray emission of DMs based on the stellar luminosity and the mass loss rate from the radiative cooling of the magnetically confined material. A main result of the XADM model is an increasing trend of X-ray emission with the spectral type and the size of the last closed field loop ($R_A$; Alfv\'en radius) for B-type stars \citep{xpod}.\\
\indent The rotational periods of some magnetic B-type stars are short enough for rotation to be dynamically significant, causing an added centrifugal component that affects the magnetically confined material. \cite{rrm} developed a rigidly rotating magnetosphere (RRM) model relying on the assumption that the magnetic field lines will be forced to co-rotate with the stellar surface. For closed field lines larger than the Kepler co-rotation radius ($R_K$; the radius at which the centrifugal acceleration is equal to the gravitational acceleration) the confined material will not fall back to the stellar surface.  Therefore, the material will accumulate creating a centrifugal magnetosphere (CM) with a disc-like structure.  Perhaps the added centrifugal acceleration on the material trapped in the CM could provide a faster velocity causing larger shock temperatures and, consequently, harder X-ray energies or higher X-ray luminosity. Another possibility to explain this over-luminosity could be an increased X-ray efficiency factor in the CM regions through lack of material fall-back as compared to the DM regions. \\
\begin{figure*}
\begin{subfigure}
\centering
\includegraphics[width=0.49\textwidth]{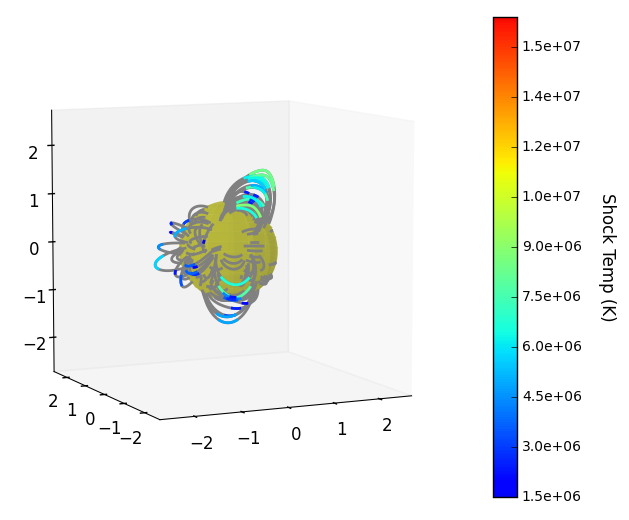}
\end{subfigure}
\begin{subfigure}
\centering
\includegraphics[width=0.49\textwidth]{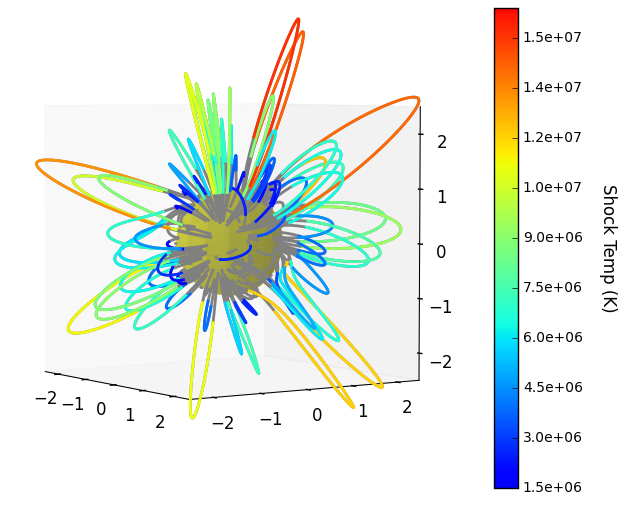}
\end{subfigure}
\caption{A 3-D representation of the field lines for the source radius $2$R$_\ast$ (left) and $5$R$_\ast$ (right) with the post shock temperatures plotted along the shock region of the individual loops. In order to reproduce the observed spectral features, there would need to be shock temperatures that are represented in red, however, for the $2$R$_\ast$ there no loops producing high enough shock temperatures. The $5$R$_\ast$ has larger closed field loops that result in higher shock temperatures. \label{fig:tausco}}
\end{figure*}
\indent \cite{xpod} performed a study of all the known magnetic OB-type stars for which modern X-ray observations exist.  In this study the observed X-ray luminosity from the Chandra X-ray Observatory and/or the XMM-Newton space telescope was compared to the predicted X-ray luminosity from the XADM model (Fig. \ref{fig:newdetect}). Although the agreement is generally good, we can see that a number of stars have observed X-ray luminosity higher than that predicted by the XADM model.  Most of the overluminous stars in Fig. \ref{fig:newdetect} are fast rotators. However, the one overluminous star that is not a fast rotator is $\tau$ Sco; this star has a complex surface field configuration, which is not common for fossil fields and could be the reason for its overluminosity.  \\ 
\section{$\tau$ Sco: a Slow Rotator with a Complex Field}
One of the overluminous stars identified by \cite{xpod} and shown in Fig. \ref{fig:newdetect} is a slow rotator observed to have a complex surface magnetic field \citep{Donati}. The surface magnetic field can be extrapolated in order to predict the three dimensional structure of the field loops above the surface. A key parameter needed for the extrapolation is the source radius, the radius corresponding to the apex of the largest closed field loop for an arbitrary field configuration, which is determined by the ratio of the magnetic field energy density to the stellar wind kinetic energy density. A smaller mass loss rate will produce a larger source radius. As the mass-loss rate is relatively unconstrained, \cite{Donati} chose a mid range mass loss rate and extrapolated the surface field out to the corresponding source radius ($2$R$_\ast$), resulting in small loops. \\ 
\indent $\tau$ Sco is an ideal case for comparing the shock temperatures of the field loops to the observed spectral features because of the high resolution grating spectra obtained by \cite{Cohen03} with the Chandra X-ray Observatory. In order to do this, the ADM formalism was adapted for an arbitrary loop configuration, shown in Fig. \ref{fig:tausco}, to determine the shock temperatures for the source radius assumed by \cite{Donati}. The X-ray spectra suggests shock temperatures $\sim 15$MK (shown in red in Fig. \ref{fig:tausco}) are needed. However, the shock temperatures from the $2$R$_\ast$ source radius in Fig. \ref{fig:tausco} (left) are insufficient, suggesting that a larger source radius is needed.
\indent To find the shock temperatures for larger source radii, a field extrapolation was performed based on the surface field map of \cite{Donati} but with source radii of $3$R$_\ast$, $4$R$_\ast$, and  $5$R$_\ast$.   Determining the shock temperatures the same way as before, Fig. \ref{fig:tausco} (right) shows that the larger source radii produces higher shock temperatures that are $\sim 15$MK. Using these results to constrain the source radius will help to determine what is causing the overluminosity of $\tau$ Sco. \\
\section{Rapid Rotators}
New X-ray spectral observations of 5 rapidly rotating B-type stars were obtained with XMM-Newton Space Telescope to add to the previous X-ray observations listed by \citet{xpod}. From these observations, three new detections were obtained, while two were undetected. The spectra of the detected stars with sufficient count rates were modeled using APEC thermal plasma models to determine the total X-ray flux and temperatures. For the two non-detections the upper limit X-ray luminosity was determined. The X-ray luminosities were added to the plot comparing the predicted and observed X-ray luminosity shown in Fig. \ref{fig:newdetect} with the detected stars shown in numbered triangles and the upper limits in the numbered rectangles.  Of the three detections, one falls in the overluminous outlier group (\#60) with the other two being reasonably close to their predicted luminosity. \\
\indent Further efforts will help determine whether the added centrifugal force provides sufficient acceleration to the confined material to lead to the high post-shock temperatures inferred observationally. A comparison of the X-ray production efficiency could also lead to an understanding of the discrepancies in the observed X-ray luminosity.\\

\end{document}